\documentclass[12pt, twoside, a4paper]{article}
\usepackage{imm17_4}

\pdfoutput=1

\usepackage[pdftex, unicode]{hyperref}
\usepackage[pdftex]{graphicx}

\hypersetup{pdftitle={Estimation of the TQ-complexity of chaotic sequences}}
\hypersetup{pdfauthor={A. V. Makarenko, Constructive Cybernetics Research Group}}
\hypersetup{pdfsubject={Nonlinear Dynamics}}
\hypersetup{pdfkeywords={Discrete-time systems, Time-series analysis, Stochastic complexity, Estimation algorithms, Chaos theory, TQ-complexity, Symbolic CTQ-analysis}}

\newcommand{\MaincltStr}{arXiv.org (Proceedings of the 1st IFAC Conference MICNON~2015)}

\newcommand{\Leftclt}{A.\,V.\,Makarenko}

\newcommand{\Rightclt}{Estimation of the TQ-complexity of chaotic sequences}

\newcommand{\UDK}{MSC: 37M10, 34C28, 11Y16, 37M25, 94A17;
\\ PACS: 02.70.-c, 05.45.-a, 05.45.Tp, 89.70.Cf, 89.75.-k.}

\begin{document}

\Mainclt 

\begin{center}
\Large{\bf Estimation of the TQ-complexity of chaotic sequences}\\[2ex]
\end{center}

\begin{center}
\large\bf{A.\,V.\,Makarenko}\supit{a,}\supit{b,}
\footnote{E-mail: avm.science@mail.ru}\\[2ex]
\end{center}

\begin{center}
\supit{a}\normalsize{Constructive Cybernetics Research Group}
\\
\normalsize{P.O.Box~560, Moscow, 101000 Russia}\\[3ex]

\supit{b}
\normalsize{Institute of Control Sciences, Russian Academy of Sciences}
\\
\normalsize{ul.~Profsoyuznaya~65, Moscow, 117977 Russia}\\[3ex]
\end{center}

{\small Received January 19, 2015;\;in final form, April 22, 2015.}
\begin{quote}\small
{\bf Abstract}. A new approach is proposed to the quantitative estimation of the complexity of multidimensional discrete sequences in terms of the shapes of their trajectories in the extended space of states. This approach is based on the study of the structural properties of sequences and is suitable for estimating the complexity of both chaotic and stochastic sequences. It is constructed on the method, proposed earlier by the author, of symbolic CTQ-analysis of multidimensional discrete sequences and mappings. The algorithm proposed manipulates not only the frequency of occurrence of symbols, but also takes into account their sequence order. An example (financial time series) is given that demonstrates the application of the tools developed.
\end{quote}

\begin{Keyworden}
Discrete-time systems, Time-series analysis, Stochastic complexity, Estimation algorithms, Chaos theory, TQ-complexity, Symbolic CTQ-analysis.
\end{Keyworden}


\setcounter{equation}{0}
\setcounter{lem}{0}
\setcounter{teo}{0}



\section{Introduction}

The notion of "complexity" of an object is one of its most important structural--information characteristics and belongs to the class of fundamental scientific concepts~\cite{bibl:book_Complexity_Science_Society_2007, bibl:book_Measurements_Complexity_LNP_1988_314}. The narrower notion of the "complexity of a dynamic process" is not an exception. This notion is related to the predictability and information capacity of processes~\cite{bibl:book_Complexity_Hierarchical_Structures_Scaling_Physics_1997}. The complexity of a dynamic process is a part of the criteria for the classification of processes into deterministic, chaotic, and stochastic ones~\cite{bibl:article_Kravtsov_SPU_1989_32}. However, along with this, the questions of the definition and calculation of the complexity of dynamic processes remain methodologically open~\cite{bibl:book_Measurements_Complexity_LNP_1988_314}.

A quantitative approach to the notion of complexity was first formulated in the statistical physics of equilibrium systems in~1877 by Ludwig Boltzmann, who introduced the concept of "entropy"~\cite{bibl:book_Leontovich_Statistical_Physics_1983}, $H = k_B\ln W$, where $W$ is the number of microstates of a system that can be implemented in the existing macroscopic state and $k_B$ is the Boltzmann constant. R.~Hartley actually extended the principles of statistical physics to the description of the states of macrosystems and gave entropy an informational meaning~\cite{bibl:article_Hartley_Bell_System_Technical_Journal_1928_7}.

This idea was further developed in the works of T.~Shannon on information theory~\cite{bibl:article_Shannon_Bell_System_Technical_Journal_1948_27}.
In these works, Shannon also introduced the notion of entropy:
\begin{equation}\label{eq:Entr_Shen}
 H=-\sum\limits_{i}p(x_i)\,\ln p(x_i),
\end{equation}
where~$p(x_i)$ is the probability distribution of independent random events~$x_i$. Shannon's entropy was generalized to dynamical systems by A.~N.~Kolmogorov and Y.~G.~Sinai in their entropy theory of dynamical systems~\cite{bibl:book_Kornfeld_Sinai_Dynamical_Systems_1985_2}.

The development of nonlinear dynamics, the theory of chaotic dynamical systems, and the theory of non-equilibrium systems required the introduction of appropriate characteristics, such as Lyapunov exponents, Kolmogorov entropy, and Klimontovich's S-parameter~\cite{bibl:book_Kuznetsov_Dynamical_Chaos_2001, bibl:book_Klimontovich_Structure_chaos_1990}. It is noteworthy that these parameters are also inherently linked to Shannon's entropy.

However, all widely used modifications of the Boltzmann--Shannon entropy measure have features that limit their applicability. See, for example~\cite{bibl:article_Bashkirov_TMP_2006_149}. Therefore, R\'enyi and Tsaliss proposed new formalisms in addition to the Boltzmann--Shannon entropy.

In the early 1980's, A.~N.~Kolmogorov proposed a fundamentally new, algorithmic, approach to the interpretation of the concept of complexity~\cite{bibl:article_Kolmogorov_RMS_1983_38}. He formalized the criterion in the language of the theory of algorithms and constructed an appropriate measure for it that has
indubitable informational advantages. However, it is very difficult to apply this measure to estimating the complexity of dynamic processes, because the computation involved and the interpretation of the results are very laborious.

Darkhovskii et al.~\cite{bibl:article_Darkhovskii_Kaplan_Shishkin_ARC_2002_63}, proposed an original approach to calculating the complexity of a scalar dynamic process. The approach is based on the idea of
information expenditures needed to approximate a process to a required degree of accuracy. The approach is conceptually similar to the algorithmic approach of A.~N.~Kolmogorov. Its limitation is that the choice of the approximating basis is arbitrary and is not substantiated.

In radio physics, one actively uses the time--frequency criterion of complexity~\cite{bibl:book_Gonorovsky_Radio_Circuits_Signals_1986}. Here the measure is given by the product of the spectral width by the duration of a dynamic process:
\begin{equation}\label{eq:ST_complex}
\Delta t\Delta \omega =4{{\left[ \int\limits_{-\infty }^{+\infty }{{{t}^{2}}}{{x}^{2}}(t)dt\int\limits_{-\infty }^{+\infty }{{{\omega }^{2}}}{{\left| S(\omega ) \right|}^{2}}d\omega  \right]}^{\,1/2}},
\quad
S(\omega )=\frac{1}{2\pi }\int\limits_{-\infty }^{+\infty }{x}(t)\,{{e}^{-\imath \,\omega \,t}}dt.
\end{equation}
This criterion does not take into account the shape of the spectrum and operates with the effective values of the spectral width and the duration of a dynamic process. All this makes the evaluation of the complexity rather conditional. Moreover, measure~(\ref{eq:ST_complex}) imposes constraints on the minimum decay rate of the functions~$x(t)$ and~$|S(\omega)|$ and has an energy rather than informational meaning.

Recently, V.~I.~Arnold has suggested an approach to the calculation of the complexity of lattice sequences of the form of~${{\mathbb{Z}}_{2}}\times \mathbb{Z}$ (sequences of~0 and~1),~\cite{bibl:article_Arnold_Funct_Analsis_Other_Math_2006_1}. The method is based on the formalization of the structure of sequences: first, a self-mapping for sequences is constructed (via cyclic difference), and then this mapping is represented as a graph; the complexity of the original sequence is determined in terms of the characteristics of this graph. A strong limitation of this method is that the complexity measure constructed cannot be transferred to~${{\mathbb{R}}^{N}}\times \mathbb{Z}$-continuum processes.

We should also mention the so-called perimetric complexity of binary images~\cite{bibl:article_Attneave_Arnoult_Psychological_Bulletin_1956_53}. In this case, an image can be treated as a two-dimensional scalar field. The strongest limitation of the method is that it can be applied only to binary images (from the class~$\mathbb{Z}_2\times\mathbb{Z}^2$).

In this article, we propose a different approach to the analysis of the complexity of chaotic sequences that is based on the study of the structural properties of the sequences in terms of the shape of their trajectories. This approach is free from most of the disadvantages of the above-mentioned approaches. It is based on the method of symbolic CTQ-analysis~\cite{bibl:article_Makarenko_TPL_2012_38_155, bibl:article_Makarenko_Comput_Math_Math_Phys_2012_52}, which is aimed at the study of multidimensional discrete sequences and mappings. The formalism of CTQ-analysis studies the properties of dynamical systems, that are important from the viewpoint of identification and control of the systems and prediction of their
evolution.

The results were first presented by the author at the XXV IUPAP Conference on Computational Physics~\cite{bibl:abstract_Makarenko_Computational_Physics_2013}. The initially proposed complexity measures dealt only with the occurrence frequencies of symbols and ignored the order of symbols. In the present paper, we remove this restriction, thus expanding the analytical capabilities of the approach to estimate the complexity of discrete sequences.

Moreover, we essentially revise the principles of the symbolic CTQ analysis: we formulate encoding rules for the symbols of the base alphabet in a rigorous and formal manner, which allows us to form a complete and self-consistent set of symbols.

All calculations and visualizations are performed using Wolfram Mathematica~9.

\section{Symbolic CTQ-analysis}

Denote a discrete dynamical system in the form of a mapping
\begin{equation}\label{eq:ds}
\mathbf{s}_{k+1} = \mathbf{f}\left(\mathbf{s}_{k},\,\mathbf{p}\right),
\end{equation}
with the properties:
 $\mathbf{s}\in\mathrm{S}\subseteq\mathbb{R}^N$, $k\in\mathrm{K}\subseteq\mathbb{Z}$,
 $\mathbf{p}\in\mathrm{P}\subseteq\mathbb{R}^M$, $n=\overline{1,\,N}$, $m=\overline{1,\,M}$.

In formula~(\ref{eq:ds}), $\mathbf{s}$ is a state variable of the
system and $\mathbf{p}$ is a vector of parameters. With
mapping~(\ref{eq:ds}), we associate its trajectory in
space~$\mathrm{S}\times\mathrm{K}$, which has the form of a
semisequence~$\{\mathbf{s}_{k}\}^K_{k=1}$, $k=\overline{1,\,K}$.

\subsection{T-alphabet}

Define the initial mapping, which encodes (in terms of the final
T-alphabet) the shape of the $n$-th component of a
sequence~$\{\mathbf{s}_{k}\}^K_{k=1}$~\cite{bibl:article_Makarenko_TPL_2012_38_155, bibl:article_Makarenko_Comput_Math_Math_Phys_2012_52}:

\begin{equation}\label{eq:mapping_TSymb}
\left\lbrace\mathbf{s}^{(n)}_{k-1},\,\mathbf{s}^{(n)}_{k},\,\mathbf{s}^{(n)}_{k+1}\right\rbrace\Rightarrow
T^{\alpha\varphi}_k|_n,
\quad
T^{\alpha\varphi}_k = \left[T^{\alpha\varphi}_k|_1,\,\ldots,\,T^{\alpha\varphi}_k|_N\right].
\end{equation}

\begin{figure}[!htb]
\begin{center}
\includegraphics[width=131mm, height=30mm]{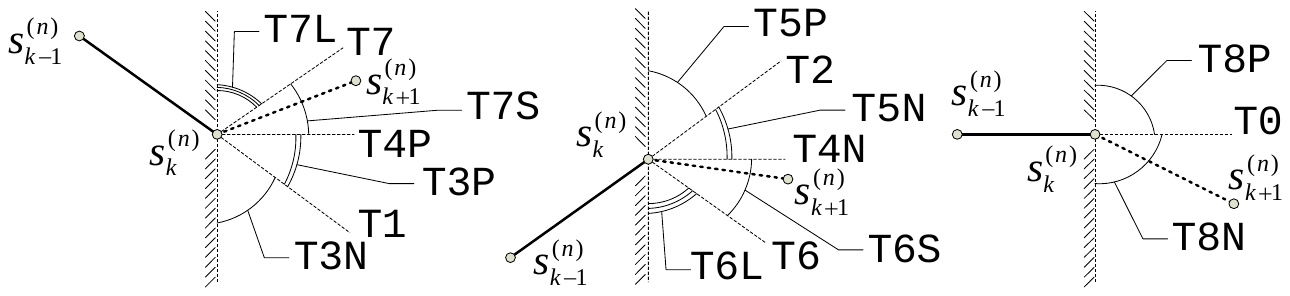}
\caption{Geometry of T-alphabet symbols.}
\label{fig:TSymb_diag}
\end{center}
\end{figure}

The graphic diagrams illustrating the geometry of the symbols~$T^{\alpha\varphi}|_n$ for the~$k$-th sample and the~$n$-th phase variable are shown in Figure~\ref{fig:TSymb_diag}.

Strictly, the mapping~(\ref{eq:mapping_TSymb}) is defined by the relations:
\begin{equation}\label{eq:TSymb}
\begin{aligned}
 &\mathtt{T0}\quad   &&\Delta s_-=\Delta s_+=0,     \\
 &\mathtt{T1}\quad   &&\Delta s_-=\Delta s_+<0,     \\
 &\mathtt{T2}\quad   &&\Delta s_-=\Delta s_+>0,     \\
 &\mathtt{T3N}\quad  &&\Delta s_-<0,\quad \Delta s_+<\Delta s_-,  \\
 &\mathtt{T3P}\quad  &&\Delta s_-<0,\quad \Delta s_+<0,\quad \Delta s_+>\Delta s_-,  \\
 &\mathtt{T4N}\quad  &&\Delta s_->0,\quad \Delta s_+=0,  \\
 &\mathtt{T4P}\quad  &&\Delta s_-<0,\quad \Delta s_+=0,  \\
 &\mathtt{T5N}\quad  &&\Delta s_->0,\quad \Delta s_+>0,\quad \Delta s_+<\Delta s_-,  \\
 &\mathtt{T5P}\quad  &&\Delta s_->0,\quad \Delta s_+>\Delta s_-,  \\
 &\mathtt{T6S}\quad  &&\Delta s_->0,\quad \Delta s_+<0,\quad \Delta s_+> -\Delta s_-,  \\
 &\mathtt{T6}\quad   &&\Delta s_-=-\Delta s_+>0,  \\
 &\mathtt{T6L}\quad  &&\Delta s_->0,\quad \Delta s_+<0,\quad \Delta s_+< -\Delta s_-,  \\
 &\mathtt{T7S}\quad  &&\Delta s_-<0,\quad \Delta s_+>0,\quad \Delta s_+< -\Delta s_-,  \\
 &\mathtt{T7}\quad   &&\Delta s_-=-\Delta s_+<0,  \\
 &\mathtt{T7L}\quad  &&\Delta s_-<0,\quad \Delta s_+>0,\quad \Delta s_+> -\Delta s_-,  \\
 &\mathtt{T8N}\quad  &&\Delta s_-=0,\quad \Delta s_+<0,  \\
 &\mathtt{T8P}\quad  &&\Delta s_-=0,\quad \Delta s_+>0.
\end{aligned}\,,
\end{equation}
here~$\Delta s_- = \mathbf{s}^{(n)}_{k}-\mathbf{s}^{(n)}_{k-1}$
and~$\Delta s_+ = \mathbf{s}^{(n)}_{k+1}-\mathbf{s}^{(n)}_{k}$.

Thus, the T-alphabet includes the following set of symbols:
\begin{multline} \label{eq:T_alphabet_o_n}
\mathrm{T}^{\alpha\varphi}_o=\{\mathtt{T0},\,\mathtt{T1},\,\mathtt{T2},\,\mathtt{T3N},
\,\mathtt{T3P},\,\mathtt{T4N},\,\mathtt{T4P},\,\mathtt{T5N},\,\mathtt{T5P},\,
\\
\mathtt{T6S},\,\mathtt{T6},\,\mathtt{T6L},\,\mathtt{T7S},\,\mathtt{T7},\,\mathtt{T7L},
\,\mathtt{T8N},\,\mathtt{T8P}\}.
\end{multline}

One can see from~(\ref{eq:T_alphabet_o_n}) that the symbol~$T^{\alpha\varphi}_k|_n$ is encoded as~$\mathtt{T}\,i$, where~$i$ is the right-hand side of the symbol codes of the alphabet~$\mathrm{T}^{\alpha\varphi}_o$. In turn, the symbol~$T^{\alpha\varphi}_k$ is encoded in terms of~$\mathtt{T}\,i_1\,\cdots\,i_N$, see~(\ref{eq:mapping_TSymb}). The full alphabet~$\mathrm{T}^{\alpha\varphi}_o|N$, which encodes the shape of the trajectory of the multidimensional sequence~$\{\mathbf{s}_k\}^K_{k=1}$, consists of~$17^N$~symbols.

\subsection{Q-alphabet}

In addition to the symbols~$T^{\alpha\varphi}_k|_n$, we introduce the symbols~$Q^{\alpha\varphi}_k|_n$:
\begin{equation}\label{eq:SymbQ}
Q^{\alpha\varphi}_k|_n\equiv T^{\alpha\varphi}_k|_n\rightarrow T^{\alpha\varphi}_{k+1}|_n,
\quad
Q^{\alpha\varphi}_k = \left[Q^{\alpha\varphi}_k|_1,\,\ldots,\,Q^{\alpha\varphi}_k|_N\right].
\end{equation}

\begin{figure}[!htb]
\begin{center}
\includegraphics[width=76mm, height=76mm]{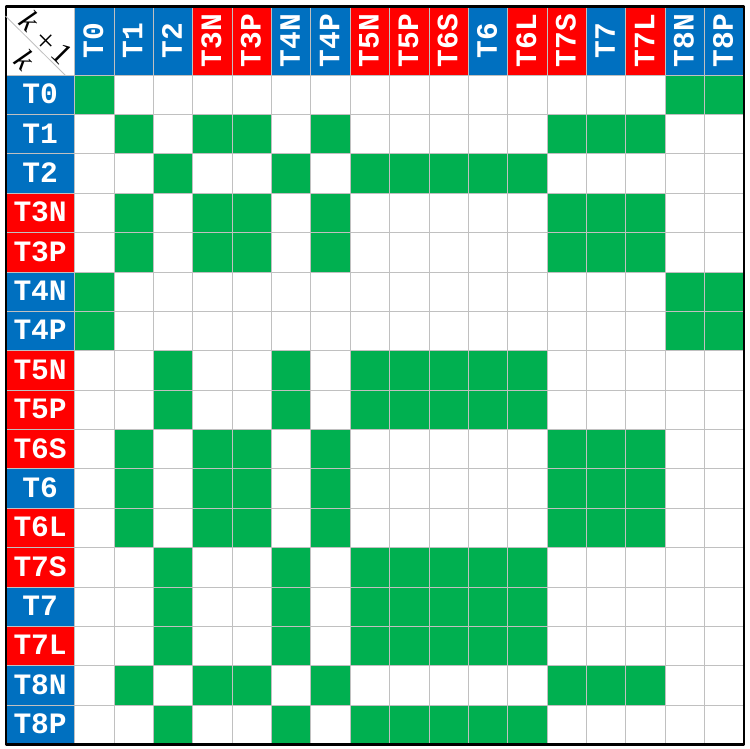}
\caption{Table of the transitions~$T^{\alpha\varphi}_k|_n\rightarrow T^{\alpha\varphi}_{k+1}|_n$; admissible transitions are shown in green.}
\label{fig:TSymb_table_trans}
\end{center}
\end{figure}

All admissible transitions constitute a set of symbols of the alphabet~$\mathrm{Q}^{\alpha\varphi}_o\ni Q^{\alpha\varphi}_k|_n$. These transitions are shown in Figure~\ref{fig:TSymb_table_trans}.

The symbol~$Q^{\alpha\varphi}_k|_n$ is encoded as~$\mathtt{Q}\,i\,j$, where~$i$ and~$j$ are the right-hand sides of the symbol codes of the alphabet~$\mathrm{T}^{\alpha\varphi}_o$ for the states $k$ and $k+1$, respectively. In turn, the symbol~$Q^{\alpha\varphi}_k$ is encoded in terms of~$\mathtt{Q}\,i_1\,\cdots\,i_N\,j_1\,\cdots\,j_N$, see~(\ref{eq:SymbQ}). The full alphabet~$\mathrm{Q}^{\alpha\varphi}_o|N$, which encodes the shape of the trajectory of the
sequence~$\{\mathbf{s}_k\}^K_{k=1}$, consists of~$107^N$~symbols (see Figure~\ref{fig:TSymb_table_trans}).

\subsection{Symbolic TQ-image of a dynamical system}

Let us introduce a finite graph
\begin{equation} \label{eq:Gamma_TQ_n}
\Gamma^{TQ}|_n=\left\langle\mathrm{V^\Gamma}|_n,\,\mathrm{E^\Gamma}|_n\right\rangle,\quad
\mathrm{V^\Gamma}|_n\subseteq\mathrm{T}^{\alpha\varphi}_o,\,
\mathrm{E^\Gamma}|_n\subseteq\mathrm{Q}^{\alpha\varphi}_o,
\end{equation}
where~$\mathrm{V}^{\Gamma}|_n$ is the vertex set and~$\mathrm{E}^{\Gamma}|_n$ is the edge set of~$\Gamma^{TQ}|_n$. According to its topology, the graph~$\Gamma^{TQ}|_n$ is a connected directed graph without multiple arcs but with loops. The graph~$\Gamma^{TQ}|_n$ is a particular symbolic TQ-image of the dynamical system with respect to its $n$-th phase variable.

The set of graphs
\begin{equation} \label{eq:Gamma_TQ}
\Gamma^{TQ}=\left[\Gamma^{TQ}|_1,\,\ldots,\,\Gamma^{TQ}|_N\right],
\end{equation}
is a complete symbolic TQ-image of the dynamical system.

Let us denote the graph~$\Gamma^{TQ}|_n$ corresponding to the full alphabets~$\mathrm{T}^{\alpha\varphi}_o$
and~$\mathrm{Q}^{\alpha\varphi}_o$ by~$\Gamma^{TQ}_o$.

The graph~(\ref{eq:Gamma_TQ_n}) can be weighted (on its vertices and edges) by the occurrence frequency of characters~$*$ in the sequence~$\{\mathbf{s}^{(n)}_{k}\}^K_{k=1}$:
\begin{equation}\label{eq:Delta_o}
\Delta^*|_n = \frac{\left| {\mathrm{M}^*|_n} \right|}{
\left|\bigcup\limits_{*}\mathrm{M}^*|_n\right|},\quad 0\leqslant\Delta^*|_n\leqslant 1,
\end{equation}
where $|\circ|$ is the cardinality of the set and $*$ is a symbol of which the multiset~$\mathrm{M}^*|_n$ consists:
\begin{subequations}\label{eq:Delta_Symb_All}
\begin{align}
&\Delta^T|_n: \mathrm{M}^*|_n \ni T^{\alpha\varphi}_k|_n:\,T^{\alpha\varphi}_k|_n\backslash\mathtt{T} = * ,\;
* \in \mathrm{T}^{\alpha\varphi}_o\backslash\mathtt{T},
                \label{eq:Delta_Symb_T} \\[3pt]
&\Delta^Q|_n: \mathrm{M}^*|_n \ni Q^{\alpha\varphi}_k|_n:\,Q^{\alpha\varphi}_k|_n\backslash\mathtt{Q} = * ,\;
* \in \mathrm{Q}^{\alpha\varphi}_o\backslash\mathtt{Q}.
                \label{eq:Delta_Symb_Q}
\end{align}
\end{subequations}

Note that the calculation of~(\ref{eq:Delta_Symb_T}) and~(\ref{eq:Delta_Symb_Q}) allows one to quantitatively assess various properties of the trajectory of the sequence~$\{s^{(n)}_{k}\}^K_{k=1}$ in the space~$\mathrm{S}^{(n)}\times\mathrm{K}$, including the Markov characteristic of the sequence~$\{T^{\alpha\varphi}_k|_n\}^K_{k=1}$~\cite{bibl:article_Bowen_AJM_1973_95}.

\section{Measurement of TQ-complexity}
\label{sect:TQ_Complexity_Measures}

The approach presented here to the calculation of the complexity of multidimensional discrete mappings and sequences is informally defined by the following statement: {\it The more complex is a dynamic process, the more complex is the shape of its trajectory in the space~$\mathrm{S}\times\mathrm{K}$}. Below, we present this statement in a formalized manner.

First, to each of the symbols~$T^{\alpha\varphi}|_n$ and~$Q^{\alpha\varphi}|_n$, we assign a numerical value of the complexity --- the so-called {\it unit complexity of a symbol}: $C^T|_n$ and~$C^Q|_n$.

The symbol~$T^{\alpha\varphi}|_n$ is composite~\cite{bibl:article_Makarenko_Comput_Math_Math_Phys_2012_52};
therefore, we first determine the unit complexity of their constituent elementary symbols:
\begin{subequations}\label{eq:Unit_TQComplexity_Symb_S}
\begin{align}
&C^\alpha|_n=
\begin{cases}
1 & S^\alpha|_n = \mathtt{Z},\\
2 & S^\alpha|_n = \mathtt{D},\,\mathtt{U}.
\end{cases},
\\[3pt]
&C^\varphi|_n=
\begin{cases}
1 & S^\varphi|_n = \mathtt{L},\\
2 & S^\varphi|_n = \mathtt{B},\\
3 & S^\varphi|_n = \mathtt{E}.
\end{cases}.
\end{align}
\end{subequations}

The unit complexity of the symbol~$T^{\alpha\varphi}|_n$ is represented as~$\mathbf{C}^T|_n=\left[ C^T_\alpha|_n,\, C^T_\varphi|_n \right]^\mathrm{T}$, where~$\cdot^\mathrm{T}$ is the transpose operator. The norm of this quantity is defined as~$C^T|_n = C^T_\alpha|_n + C^T_\varphi|_n -1$.

The values of the components~$\mathbf{C}^T|_n$ are given in table~\ref{tbl:Unit_TQComplexity_TSymb}.
\begin{table}[h!tb]
\renewcommand{\arraystretch}{1.2}
\begin{center}
\caption{Unit complexities of the symbols~$T^{\alpha\varphi}|_n$,
($*=\mathtt{N},\,\mathtt{P}$; $\circ=\mathtt{S},\,\mathtt{L}$).}
\label{tbl:Unit_TQComplexity_TSymb}
\begin{tabular}{ccccccc}
$T^{\alpha\varphi}|_n$
&$\mathtt{T0}$
&$\mathtt{T1}$, $\mathtt{T2}$
&$\mathtt{T4}*$, $\mathtt{T8}*$
&$\mathtt{T3}*$, $\mathtt{T5}*$
&$\mathtt{T6}$, $\mathtt{T7}$
&$\mathtt{T6\circ}$, $\mathtt{T7\circ}$\\ \hline
$C^T_\alpha|_n$  & 1 & 2 & 3 & 4 & 2 & 4 \\
$C^T_\varphi|_n$ & 1 & 1 & 2 & 2 & 3 & 3 \\ \hline
$C^T|_n$         & 1 & 2 & 4 & 5 & 4 & 6 \\
\end{tabular}
\end{center}
\end{table}
\\Note that the table is compiled on the following key principle: repeated symbols (subsequences) do not increase the complexity of the sequence, since they do not carry new information.

\begin{figure}[h!tb]
\begin{center}
\includegraphics[width=83mm, height=55mm]{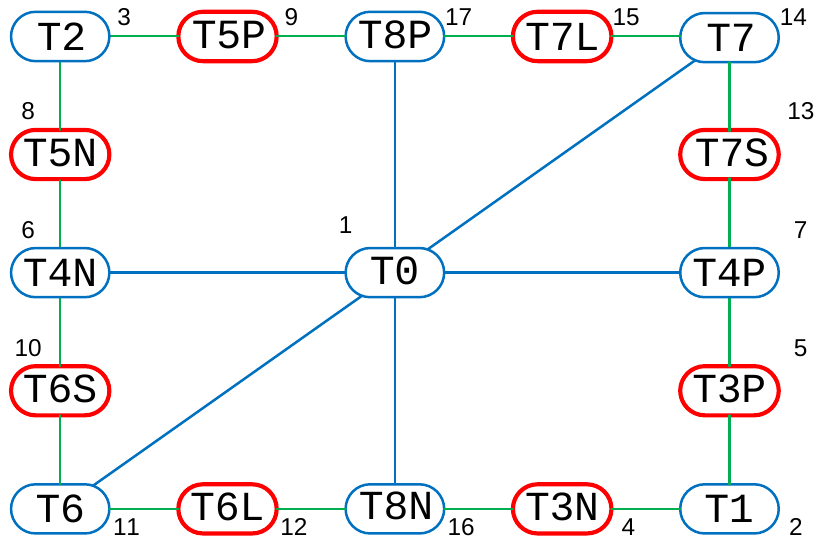}
\caption{
The graph~$\mathrm{D^\mathrm{1p}_T}$ corresponds to transitions
between the symbols~$T^{\alpha\varphi}|_n$ for the $k$-th sample
of the subsequence
$\left\{s^{(n)}_{k-1},\,s^{(n)}_k,\,s^{(n)}_{k+1}\right\}$ under
its various continuous one-point deformations.
See~\cite{bibl:abstract_Makarenko_Analysis_Singularities_2012}.}
\label{fig:Graph_D_T_1p}
\end{center}
\end{figure}

Define the unit complexity of the symbol~$Q^{\alpha\varphi}_k|_n$ in terms of the distance between~$T^{\alpha\varphi}_k|_n$ and~$T^{\alpha\varphi}_{k+1}|_n$:
\begin{equation} \label{eq:UnitTQComplexity_QSymb}
C^Q|_n = \mathrm{d_T}\left(T^{\alpha\varphi}_k|_n,\,T^{\alpha\varphi}_{k+1}|_n\right)+1.
\end{equation}
The measure~$\mathrm{d_T}\left(\cdot,\,\cdot\right)$ is the number of edges on the shortest path between two vertices in the graph~$\mathrm{D^\mathrm{1p}_T}$ (see Figure~\ref{fig:Graph_D_T_1p}).

In contrast to the earlier paper~\cite{bibl:abstract_Makarenko_Computational_Physics_2013}, we use a scheme of one-point deformation of the subsequence~$\left\{s^{(n)}_{k-1},\,s^{(n)}_k,\,s^{(n)}_{k+1}\right\}$
when constructing the graph~$\mathrm{D^\mathrm{1p}_T}$. This construction is closer to the classical Levenshtein distance~\cite{bibl:book_Levenshtein_SPD_1966}, with the following edit transcript: {\bf R}eplace and {\bf M}atch~\cite{bibl:book_Algorithms_Strings_Trees_Sequences_1997}. Thus, this modification makes the specific complexity of the symbols~$Q^{\alpha\varphi}_k|_n$ more strictly defined. Moreover, the range of values of~$C^Q|_n$ becomes balanced with resect to the range of~$C^T|_n$.

As already pointed in the introduction, the measures of complexity (based on symbolic CTQ-analysis) proposed earlier by the author ignored the order of symbols~\cite{bibl:abstract_Makarenko_Computational_Physics_2013}. This is a significant restriction. Indeed, consider two test sequences (letter~$\mathtt{T}$ in notation of the symbols of the T-alphabet is omitted):
\begin{subequations}\label{eq:test_seq}
\begin{align}
\mathtt{7S}\,\mathtt{5P}\,\mathtt{6L}\,\mathtt{7S}\,\mathtt{5P}\,\mathtt{6L}\,
\mathtt{7S}\,\mathtt{5P}\,\mathtt{6L}\,\mathtt{7S}\,\mathtt{5P}\,\mathtt{6L}\,
\mathtt{7S}\,\mathtt{5P}\,\mathtt{6L}\,\mathtt{7S}\,\ldots,
                \label{eq:test_seq_period} \\[3pt]
\mathtt{7S}\,\mathtt{6L}\,\mathtt{7S}\,\mathtt{5P}\,\mathtt{5P}\,\mathtt{5P}\,
\mathtt{6L}\,\mathtt{7S}\,\mathtt{5P}\,\mathtt{6L}\,\mathtt{7S}\,\mathtt{5P}\,
\mathtt{6L}\,\mathtt{7S}\,\mathtt{6L}\,\mathtt{7S}\,\ldots.
                \label{eq:test_seq_nonperiod}
\end{align}
\end{subequations}
Intuitively and logically, the sequence~(\ref{eq:test_seq_period}) is simpler. It shows clear periodicity, which can easily be continued. The second sequence~(\ref{eq:test_seq_nonperiod}) is objectively more complicated. It is not so easy to continue (the principle of its generation is unclear). At the same time, both sequences~(\ref{eq:test_seq_period}) and~(\ref{eq:test_seq_nonperiod}) have the same set of symbols and
differ only in their sequence order. Thus, to distinguish sequences (periodic sequence from random and chaotic ones), we introduce a reduction procedure.

\begin{figure}[h!tb]
\begin{center}
\includegraphics[width=132mm, height=57mm]{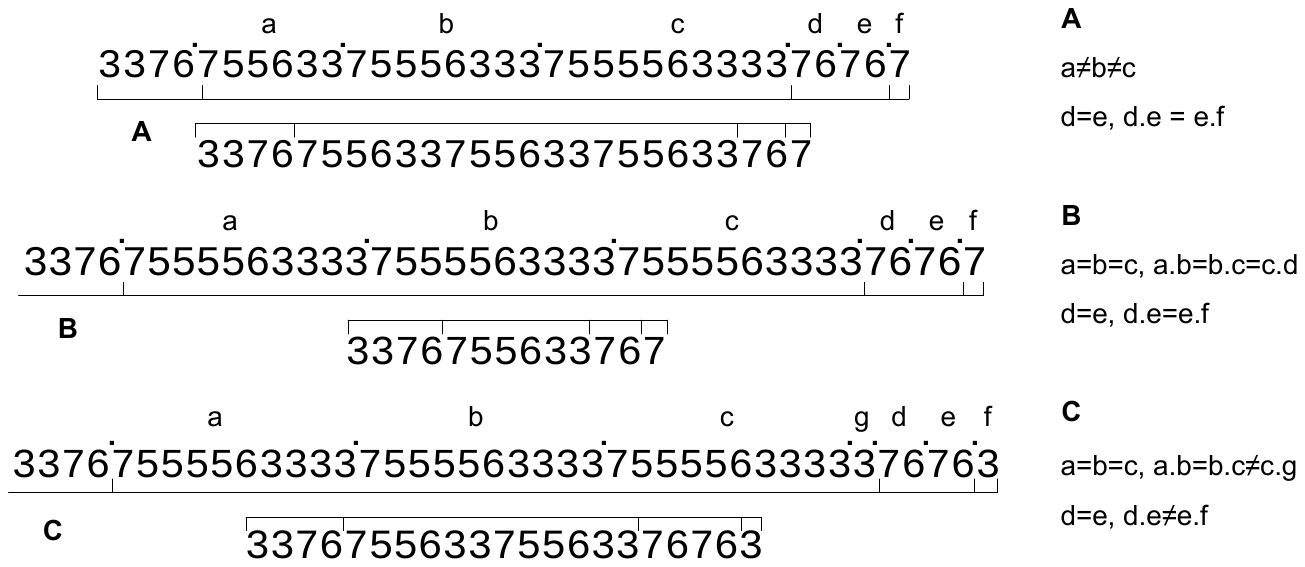}
\caption{Illustration of the reduction rule of the sequences~$\{T^{\alpha\varphi}_k|_n\}^K_{k=1}$
(letter~$\mathtt{T}$ in notation of the symbols of the T-alphabet is omitted); the sign~"." denotes the boundary between segments (in fact, it is a transition~$T^{\alpha\varphi}_k|_n\to T^{\alpha\varphi}_{k+1}|_n$, i.e., the
symbol~$Q^{\alpha\varphi}|_n$).} \label{fig:DropSeqRules}
\end{center}
\end{figure}

When calculating the TQ-complexity of the sequence $\{T^{\alpha\varphi}_k|_n\}^K_{k=1}$, one should first
reduce it; i.e., one should remove repeated subsequences, because they do not carry new information. The reduction rule is illustrated in Figure~\ref{fig:DropSeqRules}.

The removal of duplicate fragments is performed starting from longer to shorter ones. This condition allows one to distinguish between periodic and quasi-periodic segments. Furthermore, the removal of identical fragments is performed so that, locally (within the deleted blocks), the set of~$T^{\alpha\varphi}|_n$ and~$Q^{\alpha\varphi}|_n$ symbols is preserved. This guarantees the invariance of the graph~$\Gamma^{TQ}|_n$. After the application of this rule, we obtain a reduced sequence~$\{T^{\alpha\varphi}_k|_n\}^{K'}_{k=1}$.

Let us introduce two measures of complexity.

{\it Degenerate measure}~$\mathbf{C}^\mathrm{d}_S = \left[C^\mathrm{d}_{ST},\,C^\mathrm{d}_{SQ}\right]^\mathrm{T}$.
This measure deals solely with the length of a reduced sequence.

Formally, the unit complexities of the symbols~$T^{\alpha\varphi}|_n$ and~$Q^{\alpha\varphi}|_n$ are
assumed to be equal to unity:
\begin{equation}\label{eq:TQComplex_Measur_d}
C^\mathrm{d}_{ST}|_n = K', \quad
C^\mathrm{d}_{SQ}|_n = K'-1.
\end{equation}

{\it Weighted measure}~$\mathbf{C}^\mathrm{w}_S = \left[C^\mathrm{w}_{ST},\,C^\mathrm{w}_{SQ}\right]^\mathrm{T}$.
This is an extension of the degenerate measure.

That is, the weighted measure also includes the unit complexities
of the symbols~$T^{\alpha\varphi}|_n$ and~$Q^{\alpha\varphi}|_n$:
\begin{equation}\label{eq:TQComplex_Measur_w}
C^\mathrm{w}_{ST}|_n =\sum\limits^{K'}_{k=1}C^T|_n\left[T^{\alpha\varphi}_k|_n\right],
\quad
C^\mathrm{w}_{SQ}|_n =\sum\limits^{K'-1}_{k=1}C^Q|_n\left[T^{\alpha\varphi}_k|_n, \,T^{\alpha\varphi}_{k+1}|_n\right].
\end{equation}

On the basis of the complexities~(\ref{eq:TQComplex_Measur_d}) and~(\ref{eq:TQComplex_Measur_w}), we can define the measure of the effective unit complexity of a sequence:
\begin{equation}\label{eq:TQComplex_Measur_eu}
C^\mathrm{eu}_{S\circ}|_n = \frac{C^\mathrm{w}_{S\circ}|_n}{C^\mathrm{d}_{S\circ}|_n},
\quad
\circ = T,\,Q.
\end{equation}

Note that, according to their design, the measures are directly related to such issues as periodic orbits, entropy of a dynamical system, etc.~\cite{bibl:article_Bowen_AJM_1973_95, bibl:book_Kornfeld_Sinai_Dynamical_Systems_1985_2}. The central element of this relation is the spectrum of the reductions~$H^\mathrm{DS}\left[L^\mathrm{DS},\,N^\mathrm{DS}\right]$. These quantities have the following meaning: $H^\mathrm{DS}$ is the number of acts of reduction, $L^\mathrm{DS}$ is the length of
a subsequence to be reduced, and $N^\mathrm{DS}$ is the number of removed fragments (it is proved that~$N^\mathrm{DS}\leqslant 2$). Note that a detailed study of this relation is the subject of our future research.

\section{Sample}

Let us demonstrate the capabilities of the tools developed by an example of the analysis of financial time series. The object of analysis is the time series of exchange rates of some world currencies (US dollar [USD], Euro [EUR], Japanese Yen [JPH], Swiss Franc [CHF], and British Pound [GBP] against Russian ruble). The analyzed period is from~01.01.1999 to~31.12.2014.

\begin{figure}[!htb]
\begin{center}
\includegraphics[width=158mm, height=60mm]{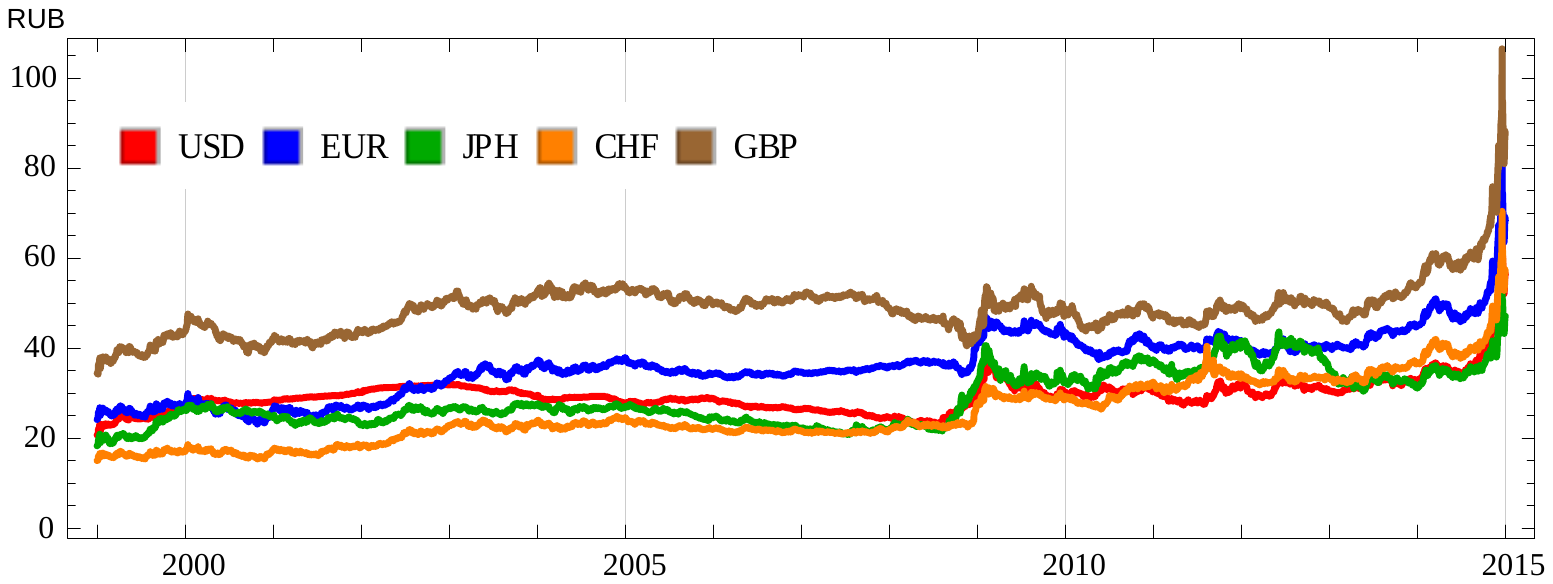}
\caption{Currency exchange rates.} \label{fig:Curency_Exch}
\end{center}
\end{figure}

Note that the analysis of the TQ-complexity has applied value in the context of research in macroeconomics and stochastic financial mathematics. The original data are taken from the official web-site of the Central Bank of Russia (Bank of Russia, exchange rates, www.cbr.ru/eng/). The length of the time series is $K=3\,985$~samples. The initial time series are shown in Figure~\ref{fig:Curency_Exch}.

\begin{figure}[!htb]
\begin{center}
\includegraphics[width=171mm, height=94mm]{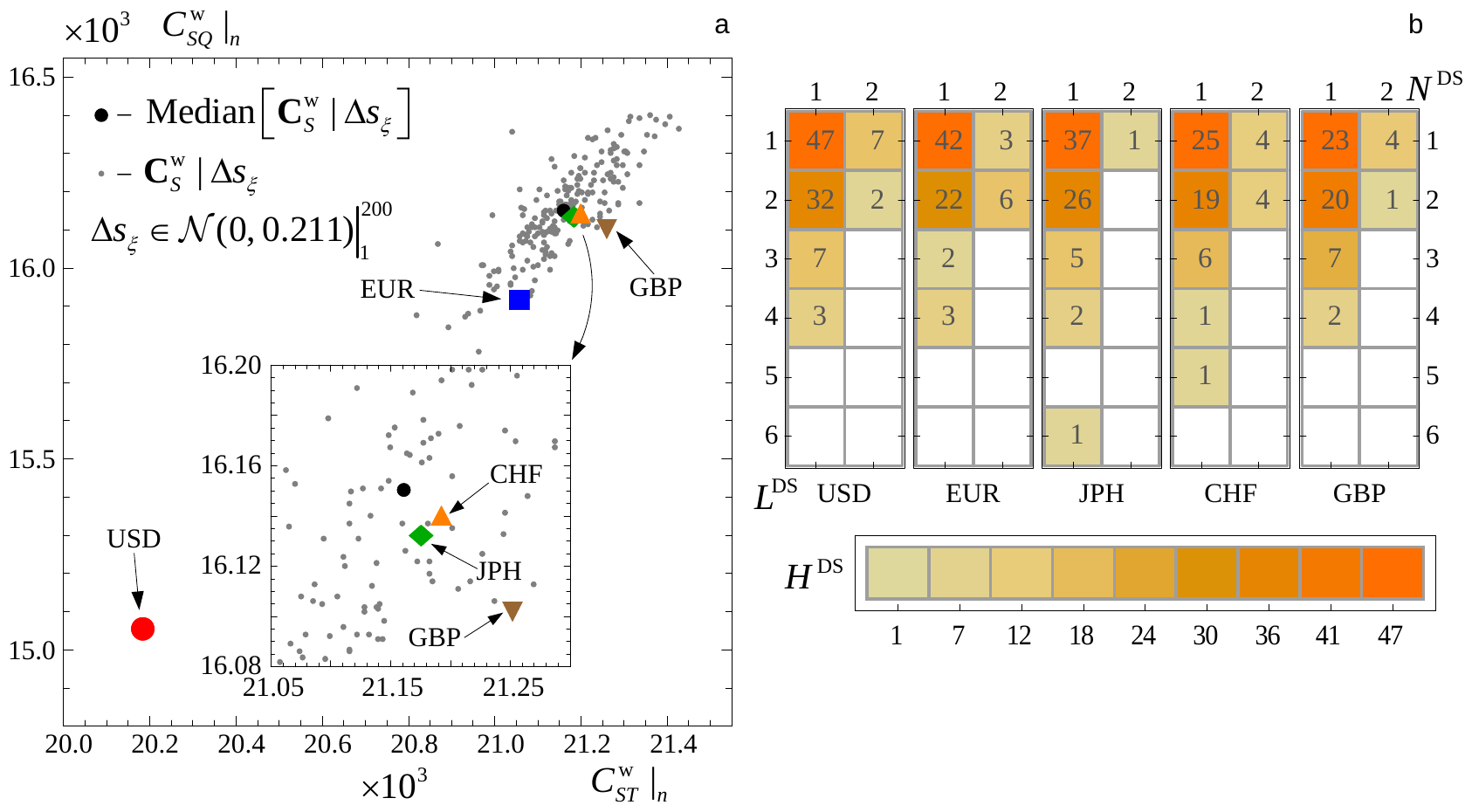}
\caption{(a) -- Estimates for the weighted measure of the TQ-complexity for the time series of currency exchange rates. (b) -- The reduction spectrum of the time series of currency
exchange rates.}
\label{fig:TQ_Complexity_HDS_Curency}
\end{center}
\end{figure}

Estimates of the weighted measure of the TQ-complexity for the
time series of currency exchange rates are shown in
Figure~\ref{fig:TQ_Complexity_HDS_Curency}a. The figure also presents
the values of~$\mathbf{C}^\mathrm{w}_S$ for the reference
stochastic sequence ($200$~realizations, length
of~$3\,985$~samples). The reference sequence has normal
distribution of discrete differences. The expectation and the
variance of the distribution are equivalent to those of the
initial sequences.

The results of the analysis imply (see Figure~\ref{fig:TQ_Complexity_HDS_Curency}a) that the pair USD/RUB
significantly differs from other currency pairs in the TQ-complexity of its time series. Moreover, the complexity of the pair USD/RUB is much lower than the complexity of the reference stochastic sequence.

From this we can draw two preliminary conclusions: (i)~the dynamics of the formation of the USD/RUB currency pair significantly differs from that of other pairs (perhaps even at the level of financial and economic mechanisms); (ii)~the time series of the USD/RUB pair is easier to predict~\cite{bibl:article_Kravtsov_SPU_1989_32}. In principle, these findings are in good agreement and complement the previous results of the author~\cite{bibl:abstract_Makarenko_AFS_2013}.

In addition, consider the reduction spectrum, which is shown in Figure~\ref{fig:TQ_Complexity_HDS_Curency}b. Figure~\ref{fig:TQ_Complexity_HDS_Curency}b shows that, predominantly, T-subsequences with a length of~$1$ and~$2$~samples are reduced. However, the JPH/RUB pair contains one single fragment with a length of~$6$~samples. It should be noted that each T-symbol comprises~$3$ consecutive samples of the initial
sequence. Information about the reduction spectrum may also be useful for the analysis of the short-term predictability of currency exchange rates.

\section{Conclusion}

In this paper, we have proposed a new approach to the quantitative evaluation of the complexity of multidimensional chaotic sequences, that is based on the study of the structural properties of sequences (in terms of the shape of their trajectories in the space~$\mathrm{S}\times\mathrm{K}$). This approach is free from most of the disadvantages of existing methods for estimating the complexity of dynamic processes. The algorithm is based on the method of symbolic CTQ-analysis. This algorithm operates not only with the frequency of occurrence of symbols, but also takes into account the sequence order of the symbols.

\section{Acknowledgments}

The author thanks Professor D.A.~Novikov, Corresponding Member of the Russian Academy of Sciences, for his attention and support.

\begin{Biblioen}

\bibitem{bibl:article_Arnold_Funct_Analsis_Other_Math_2006_1}
V.I.~Arnold.
\newblock Complexity of finite sequences of zeros and ones and geometry of finite spaces of
functions.
\newblock \emph{Funct. Analsis and Other Math.}, 1:\penalty0 1--15, 2006.

\bibitem{bibl:article_Attneave_Arnoult_Psychological_Bulletin_1956_53}
F.~Attneave, and M.D.~Arnoult.
\newblock The quantitative study of shape and pattern perception.
\newblock \emph{Psychological Bulletin}, 53:\penalty0 452--471, 1956.

\bibitem{bibl:book_Complexity_Hierarchical_Structures_Scaling_Physics_1997}
R.~Badii, and A.~Politi.
\newblock \emph{Complexity: Hierarchical Structures and Scaling in Physics}.
\newblock Cambridge University Press, Cambridge, 1997.

\bibitem{bibl:article_Bashkirov_TMP_2006_149}
A.G.~Bashkirov.
\newblock Renyi entropy as a statistical entropy for complex systems.
\newblock \emph{Theor. and Math. Phys.}, 149:\penalty0 1559--1573, 2006.

\bibitem{bibl:book_Complexity_Science_Society_2007}
J.~Bogg, and R.~Geyer, editors.
\newblock \emph{Complexity, Science and Society}.
\newblock Radcliffe Publishing, Oxford, 2007.

\bibitem{bibl:article_Bowen_AJM_1973_95}
R.~Bowen.
\newblock Symbolic dynamics for hyperbolic flows.
\newblock \emph{Amer. J. Math.}, 95:\penalty0 429--459, 1973.

\bibitem{bibl:article_Darkhovskii_Kaplan_Shishkin_ARC_2002_63}
B.S.~Darkhovskii, A.Y.~Kaplan, and S.L.~Shishkin.
\newblock On an Approach to the Estimation of the Complexity of Curves (Using as an Example an Electroencephalogram of a Human Being).
\newblock \emph{Automat. and Rem. Contr.}, 63:\penalty0 468--474, 2002.

\bibitem{bibl:book_Gonorovsky_Radio_Circuits_Signals_1986}
I.S.~Gonorovsky.
\newblock \emph{Radio Circuits and Signals}.
\newblock Radio and Communications, Moscow, 1986.

\bibitem{bibl:book_Algorithms_Strings_Trees_Sequences_1997}
D.~Gusfield.
\newblock \emph{Algorithms on Strings, Trees and Sequences}.
\newblock Cambridge University Press, Cambridge, 1997.

\bibitem{bibl:article_Hartley_Bell_System_Technical_Journal_1928_7}
R.V.~Hartley.
\newblock Transmission of Information.
\newblock \emph{Bell System Technical Journal}, 7:\penalty0 535--563, 1928.

\bibitem{bibl:book_Klimontovich_Structure_chaos_1990}
J.L.~Klimontovich.
\newblock \emph{Turbulent motion and the structure of chaos}.
\newblock Nauka, Moscow, 1990.

\bibitem{bibl:article_Kolmogorov_RMS_1983_38}
A.N.~Kolmogorov.
\newblock Combinatorial foundations of information theory and the calculus of probabilities.
\newblock \emph{Russian Mathematical Surveys}, 38:\penalty0 29--40, 1983.

\bibitem{bibl:book_Kornfeld_Sinai_Dynamical_Systems_1985_2}
I.P.~Kornfeld, and Y.G.~Sinai.
\newblock General ergodic theory of transformation groups with invariant measure. Chapter~3. Entropy theory of dynamical systems.
\newblock \emph{Dynamical systems–-2, Itogi Nauki i Tekhniki. Ser. Sovrem. Probl. Mat. Fund. Napr.}, volume~2, pages~44--70. VINITI, Moscow, 1985.

\bibitem{bibl:article_Kravtsov_SPU_1989_32}
Y.A.~Kravtsov.
\newblock Randomness, determinateness, and predictability.
\newblock \emph{Sov. Phys. Usp.}, 32:\penalty0 434--449, 1989.

\bibitem{bibl:book_Kuznetsov_Dynamical_Chaos_2001}
S.P.~Kuznetsov.
\newblock \emph{Dynamical Chaos}.
\newblock Publishing House of Sci. Lit., Moscow, 2001.

\bibitem{bibl:book_Leontovich_Statistical_Physics_1983}
M.A.~Leontovich.
\newblock \emph{Introduction to Thermodynamics. Statistical Physics}.
\newblock Nauka, Moscow, 1983.

\bibitem{bibl:book_Levenshtein_SPD_1966}
V.I.~Levenshtein.
\newblock Binary codes capable of correcting deletions, insertions, and reversals.
\newblock \emph{Sov. Phys. Doklady}, 10:\penalty0 707--710, 1966.

\bibitem{bibl:article_Makarenko_TPL_2012_38_155}
A.V.~Makarenko.
\newblock Structure of synchronized chaos studied by symbolic analysis in velocity-curvature space.
\newblock \emph{Tech. Phys. Let.}, 38:\penalty0 155--159, 2012; arXiv:1203.4214.

\bibitem{bibl:article_Makarenko_Comput_Math_Math_Phys_2012_52}
A.V.~Makarenko.
\newblock Multidimensional Dynamic Processes Studied by Symbolic Analysis in Velocity-Curvature Space.
\newblock \emph{Comput. Math. and Math. Phys.}, 52:\penalty0 1017--1028, 2012.

\bibitem{bibl:abstract_Makarenko_Analysis_Singularities_2012}
A.V.~Makarenko.
\newblock Spacing between symbols T-alphabet and properties of discrete dynamical systems.
\newblock \emph{Book of Abstracts on Int. conf. Analysis and Singularities}, pages~78--79.
Steklov Math. Inst., Moscow, 2012.

\bibitem{bibl:abstract_Makarenko_Computational_Physics_2013}
A.V.~Makarenko.
\newblock Estimation complexity of chaotic oscillations in aspect of the shape of their trajectories.
\newblock \emph{Book of Abstracts on XXV IUPAP Conf. on Computational Physics}, pages~52--53.
Department of Phys. Sci. of RAS, Moscow, 2013.

\bibitem{bibl:abstract_Makarenko_AFS_2013}
A.V.~Makarenko.
\newblock Symbolic CTQ-analysis -- a new method for studying of financial indicators.
\newblock \emph{Book of Abstracts Int. Conf. on Advanced Finance and Stochastics}, pages~63--64.
Steklov Math. Inst., Moscow, 2013.

\bibitem{bibl:book_Measurements_Complexity_LNP_1988_314}
L.~Petiti, and A.~Vulpiari, editors.
\newblock Measurements of Complexity.
\newblock \emph{Lecture Notes in Physics}, volume~314. Academic Press, New York, 3rd edition, 1988.

\bibitem{bibl:article_Shannon_Bell_System_Technical_Journal_1948_27}
C.E.~Shannon.
\newblock A Mathematical Theory of Communication.
\newblock \emph{Bell System Technical Journal}, 27:\penalty0 623--656, 1948.

\end{Biblioen}


\noindent
\\\textsf{\textbf{Andrey V. Makarenko} -- was born in~1977, since~2002 -- Ph.~D. of Cybernetics. Founder and leader of the Research \& Development group "Constructive Cybernetics". Author and coauthor of more than 60~scientific articles and reports. Member~IEEE (IEEE Signal Processing Society Membership; IEEE Computational Intelligence Society Membership). Research interests: Analysis of the structure dynamic processes, predictability; Detection, classification and diagnosis is not fully observed objects (patterns); Synchronization and self-organization in nonlinear and chaotic systems; System analysis and math.~modeling of economic, financial, social and bio-physical systems and processes; Convergence of Data~Science, Nonlinear~Dynamics, and~Network-Centric.}

\end{document}